\documentclass[useAMS,usenatbib]{mn2e}
\usepackage{amsmath}
\usepackage{amssymb}

\title[On Einstein clusters as galactic dark matter halos]{On Einstein clusters as galactic dark matter halos}

\author[C. G. B\"ohmer and T. Harko]{C. G. B\"ohmer$^{1}$\thanks{E-mail:
christian.boehmer@port.ac.uk; harko@hkucc.hku.hk} and T. Harko$^{2}$\\
$^{1}$Institute of Cosmology \& Gravitation, University of Portsmouth, Portsmouth PO1 2EG, UK\\
$^{2}$Department of Physics and Center for Theoretical and Computational Physics, The University of Hong Kong,\\ Pok Fu Lam Road, Hong Kong}
\begin{document}

\date{}

\pagerange{\pageref{firstpage}--\pageref{lastpage}} \pubyear{2007}

\maketitle

\label{firstpage}

\begin{abstract}
We consider global and gravitational lensing properties of the recently
suggested Einstein clusters of WIMPs as galactic dark matter halos. 
Being tangential pressure dominated, Einstein clusters are strongly 
anisotropic systems which can describe any galactic rotation curve
by specifying the anisotropy. Due to this property, Einstein clusters may 
be considered as dark matter candidates. We analyse the stability of the 
Einstein clusters against both radial and non-radial pulsations, and we show that
the Einstein clusters are dynamically stable. With the use of the Buchdahl type 
inequalities for anisotropic bodies, we derive upper limits on the velocity 
of the particles defining the cluster. These limits are consistent with 
those obtained from stability considerations. The study of light deflection
shows that the gravitational lensing effect is slightly smaller for the
Einstein clusters, as compared to the singular isothermal density sphere 
model for dark matter. Therefore lensing observations may discriminate, at 
least in principle, between Einstein cluster and other dark matter models.
\end{abstract}


\section{Introduction}
\label{sec1}

In 1939 Einstein~\citep{ein} presented a model of a thick spherical shell
composed of test particles of equal mass, each moving in a circular geodesic
orbit in the field of all the other. Such a system has spherical symmetry
and may be used to model globular clusters of stars, since it gives a good
approximation to the average density of matter and the average value of the
gravitational field.

Einstein clusters have been studied
extensively by~\cite{Gi54,Ho73,BaSo81,Co1,Co2}, and the main physical
properties of the cluster have been derived. In particular, the
energy-momentum tensor components $T_{i}^{j}$ for the cluster have
been obtained in several representations. Due to the spherical
symmetry, the only non-vanishing components of $T_{i}^{j}$ are
$T_{t}^{t}=\rho^{\rm (eff)}$, $T_{r}^{r}= -p_{r}^{\rm (eff)}$ and
$T_{\theta }^{\theta }=T_{\varphi }^{\varphi }=-p_{\perp}^{\rm (eff)}$, 
where $\rho^{\rm (eff)}$ is the effective energy-density
of the cluster and $p_{r}^{\rm (eff)}$ and $p_{\perp }^{\rm
(eff)}$ are the radial and tangential pressures, respectively.
Since the junction conditions require $p_{r}^{\rm (eff)}$ to be
continuous across the boundary of of each layer of the shell, it
follows that for the Einstein cluster $p_{r}^{\rm (eff)}=0$. The
non-vanishing energy-momentum components can be expressed in terms
of the velocity of the particles in the cluster and the density
only. It can also be interpreted as giving rise to rotation
without introducing global angular momentum.

Therefore, the Einstein clusters are examples of spherically symmetric 
systems with an anisotropic energy-momentum, for which the radial 
pressure is different from the tangential one, 
$p_{r}^{\rm (eff)}\neq p_{\perp}^{\rm (eff)}$. Anisotropic spherically symmetric matter
distributions have been extensively studied in the past, and
several important physical characteristics of these systems have
been obtained. In particular, several generalisations of the
Buchdahl bound on the mass-radius ratio, the redshift and the
value of the anisotropy parameter have been obtained, also
in the presence of the cosmological constant by~\cite{BoHa06}.
Similarly to the case of isotropic systems~\citep{BoHa05}, the
presence of the cosmological constant also determines the
existence of a minimal mass for anisotropic matter distributions.

An interesting physical application of the Einstein clusters has
recently been suggested by~\cite{La06}. The mass of the Einstein
cluster is proportional to the square of the tangential velocity
and to the distance to the centre. This behaviour is very similar
to that observed in the case of test particles in stable circular
orbits around the galactic centre. In most galaxies, neutral
hydrogen clouds are observed at large distances from the centre,
much beyond the extent of the luminous matter, see~\cite{Bi87,Pe}. Since
these clouds move in circular orbits with velocity $V(r)$, the
orbits are maintained by the balance between the centrifugal
acceleration $V(r)^2/r$ and the gravitational attraction force
$GM(r)/r^2$ of the total mass $M(r)$ contained within the orbit.
This allows us to express the mass profile of the galaxy in the
form $M(r)=rV^2/G$. This mass profile can be obtained rigorously
for the Einstein clusters. Observations show that the rotational
velocities increase near the centre of the galaxy and then remain
nearly constant at a value of $V_0\sim 200$ km/s~\citep{Bi87,Pe}.
Consequently, the mass within a distance $r$ from the centre of
the galaxy increases linearly with $r$, even at large distances
where very little luminous matter can be detected.

It is the purpose of the present analysis to show that the Einstein
cluster may indeed serve as a physically acceptable dark matter
model~\citep{La06}, which satisfies all energy conditions, and is
also in agreement with the generic results on anisotropic matter
distributions, obtained in~\cite{BoHa06}. As a first step we
obtain the energy-momentum tensor for the Einstein cluster, and
derive the mass-velocity relation in the presence of a
cosmological constant. Next we show that the Einstein cluster is
dynamically stable against radial and non-radial perturbations.
Several Buchdahl type physical bounds for
the Einstein cluster are also considered, leading to some upper
bounds for the velocity of the particles in the cluster and to the
estimation of its minimum mass. 

An observational possibility of testing the idea of the dark matter as 
an Einstein cluster is the study of the light deflection and lensing. 
We derive the deflection angle of light in an Einstein cluster, 
and compare its value with the deflection angle in the standard
isothermal sphere dark matter model. It is pointed out that both
models can in principle be distinguished by observation.

The present paper is organised as follows. The energy momentum
tensor for the Einstein cluster is derived in Section~\ref{sec2}. 
The stability of the Einstein cluster is considered in Section~\ref{sec3}.
General bounds on the velocity and mass of the cluster are obtained in
Section~\ref{sec4}, and the lensing effect is considered in Section~\ref{sec5}. 
We discuss and conclude our results in the final Section~\ref{sec6}. 
Throughout this paper we use a system of units so that $c=1$.

\section{The Einstein cluster model}
\label{sec2}

We consider a system consisting of particles of rest mass $m$, each
describing a circular orbit about the centre $O$ of the cluster, in the
presence of a cosmological constant $\Lambda $. The effect of the collision
between particles is neglected. Since the system is static and has spherical
symmetry, the line element is given by
\begin{align}
      ds^{2}=e^{\nu } dt^{2}-e^{\lambda }dr^{2}-r^{2}
      (d\theta^{2}+\sin^{2}\negmedspace\theta d\varphi^{2}),
      \label{line}
\end{align}
where we have denoted the coordinates as $x^{0}=t$, $x^{1}=r$,
$x^{2}=\theta$ and $x^{3}=\varphi$, respectively, $\nu $ and
$\lambda $ are functions of $r$ only, and $r\leq R$, with $r=R$
being the boundary of the cluster. This metric is joined
continuously at $r=R$ with the standard Schwarzschild - de Sitter
metric, having the mass parameter $M$, which contains the mass of
the galactic core and the mass of the dark matter halo.

The energy-momentum of the particles in the cluster is given by
\begin{align}
      T_{i}^{j}=\rho \left\langle g_{ik}\frac{dx^{j}}{ds}
      \frac{dx^{k}}{ds}\right\rangle,  
      \label{en}
\end{align}
where $\rho$ is the energy density and $\langle \rangle$ denotes
the mean value of the energy and momentum of a particle for all
the particles in the volume element around an arbitrary point $P$.
In the following we consider the particles in the cluster moving
in the plane $\varphi = {\rm constant}$. We define the components
$v^{\alpha}$ of the three-velocity of the particles, measured in
terms of the proper time, that is, by an observer located at the
given point, by~\cite{LaLi}
\begin{align}
      v^{\alpha}=\frac{dx^{\alpha}}{\sqrt{g_{tt}}dx^{t}}.
\end{align}
The square of the velocity is given by $V^{2}=v_{\alpha }v^{\alpha
}=-g_{\alpha \beta }v^{\alpha }v^{\beta }$. Since the particles
are in circular motion around the centre $O$ of the cluster, there
is only one non-vanishing component of the three-velocity,
\begin{align}
      v^{2}=v^{\theta }=e^{-\nu /2}\frac{d\theta }{dt},
\end{align}
with the magnitude of the velocity having the value
\begin{align}
      V^{2}=r^{2}e^{-\nu}\left(\frac{d\theta}{dt}\right)^{2}.
\end{align}
From Eq.~(\ref{line}) we immediately obtain
\begin{align}
      e^{\nu}\left(\frac{dt}{ds}\right)^{2} = (1-V^{2})^{-1}.
\end{align}
Then, by taking the mean value in Eq.~(\ref{en}) we obtain the components of
the energy-momentum tensor of the Einstein cluster as
\begin{align}
      T_{t}^{t} & =\rho^{\rm (eff)} = \rho (1-V^{2})^{-1}, \quad
      T_{r}^{r}=-p_{r}^{\rm (eff)}=0, \\
      T_{\theta}^{\theta} & = T_{\varphi}^{\varphi} = -p_{\perp }^{\rm (eff)}=
      -\frac{1}{2}\rho V^{2}(1-V^{2})^{-1}.
\end{align}

Anisotropic matter distributions may be described in terms of the anisotropy
parameter $\Delta =p_{\perp}^{\rm (eff)}-p_{r}^{\rm (eff)}$, which for the
Einstein cluster takes the form
\begin{align}
      \Delta  = p_{\perp }^{\rm (eff)} = \frac{1}{2}\rho V^{2} (1-V^{2})^{-1}.
      \label{delta}
\end{align}
If $\Delta >0,\forall r\neq 0$, as is the case for the Einstein cluster,
the body is tangential pressure dominated.

In the case of an anisotropic system with vanishing radial pressure the
condition of the conservation of the energy-momentum tensor, $\nabla_{j} T^{ij}=0$
gives
\begin{align}
      \nu'=\frac{4\Delta}{\rho^{\rm (eff)}r},
\end{align}
where the prime denotes differentiation with respect to $r$,
so that for the Einstein clusters we obtain
\begin{align}
      \label{nu}
      \nu'=\frac{2}{r} V^{2}.
\end{align}

In static and spherically symmetric spacetimes the field equations
reduce to three independent equations, which imply conservation of
energy-momentum. Since in the present discussion it is convenient
to use the conservation equation directly, it suffices to consider
the first two Einstein field equations, which are
\begin{align}
      -e^{-\lambda }\left(\frac{1}{r^{2}}-\frac{\lambda'}{r}\right) +
       \frac{1}{r^{2}} &= 8\pi G \rho (1-V^{2})^{-1} + \Lambda,
      \label{f1} \\
      e^{-\lambda }\left(\frac{\nu'}{r}+\frac{1}{r^{2}}\right) -
      \frac{1}{r^{2}} &= -\Lambda,
      \label{f2}
\end{align}
respectively.

The latter Eq.~(\ref{f2}) leads to
\begin{align}
      e^{-\lambda } = (1-\Lambda r^{2})(2V^{2}+1)^{-1},
      \label{l1}
\end{align}
whereas from Eq.~(\ref{f1}) we obtain
\begin{align}
      e^{-\lambda } = 1-\frac{2 G m(r)}{r}-\frac{\Lambda }{3}r^{2} ,
      \label{l2}
\end{align}
where we have defined the mass inside the radius $r$ by
\begin{align}
      m(r) = 4\pi \int_{0}^{r} \rho (1-V^{2})^{-1} r'^{2} dr'.
\end{align}

The metric function $\lambda$ can be eliminated from Eqs.~(\ref{l1}) and~(\ref{l2}) to
obtain the explicit dependence of $m(r)$ on $V^{2}$ and $r$ as
\begin{align}
      \frac{2 G m(r)}{r} = \frac{2V^{2}+\Lambda r^{2}}{1+2V^{2}} - \frac{\Lambda}{3}r^{2}.
      \label{mass}
\end{align}
Hence, the explicit dependence of the velocity on the mass of the cluster is
\begin{align}
      V^{2} = \left(\frac{G m(r)}{r} - \frac{\Lambda }{3}r^{2}\right)
      \left(1-\frac{2 G m(r)}{r} - \frac{\Lambda }{3}r^{2}\right)^{-1}.
      \label{vel}
\end{align}

For the case of low rotational velocities $V^{2} \ll 1$ (the Newtonian limit)
the expression for the tangential velocity becomes
\begin{align}
      V^{2} = \frac{G m(r)}{r} - \frac{\Lambda }{3}r^{2}.
      \label{vel2}
\end{align}
In the same approximation the metric tensor coefficient $e^\nu$ is given by
\begin{align}
      e^{\nu(r)}=\exp \left[2 \int \left(\frac{Gm(r')}{r'^{2}} -
      \frac{\Lambda}{3}r'\right) dr' \right].
\end{align}

In the Einstein cluster interpretation of the dark matter, the tangential
velocity tends to a constant value for large radii
$\lim_{r\rightarrow \infty}V(r)=V_{0}={\rm constant.}$ Therefore in this
region the density of the Einstein cluster is given by
\begin{align}
      \rho(r) = \frac{V_{0}^{2}}{4\pi G r^{2}} + \frac{\Lambda}{4\pi G}.
      \label{densit}
\end{align}

\section{Stability of the Einstein clusters against radial and non-radial perturbations}
\label{sec3}

The issue of stability is of great importance in the study of
Newtonian and general relativistic models of self-gravitating
objects. Its relevance becomes evident if we recall that any astrophysical 
model is physically uninteresting if it is unstable against small
perturbations, and different degrees of stability/instability will
lead to different patterns of evolution in the collapse of self
gravitating objects. Therefore, as realistic dark matter models
the Einstein clusters must be dynamically stable.

The stability of astrophysical objects has been the subject of many 
investigations. Most studies, however, assume that the matter is 
described by an isotropic perfect fluid. On the
other hand one expects that the anisotropy will change the evolution of 
self-gravitating systems~\cite{HK76,her97}.

In the present Section we analyse the stability of the Einstein
clusters under small radial and non-radial perturbations. In both
cases we neglect the effect of the cosmological constant since its
effects on the dynamical stability small for the present 
model, see e.g.~\cite{BoHa05a,Hl07}.

\subsection{Stability against radial perturbations}

We assume that initially the equilibrium configuration of the
Einstein cluster is spherically symmetric, and that the
perturbations preserve this symmetry. Under these perturbations 
only radial motions will ensue. 
Hence the metric of the spacetime is still given by Eq.~(\ref{line}), but 
with $\nu$ and $\lambda$ functions of both $t$ and $r$, 
$\nu =\nu(t,r)$, $\lambda =\lambda(t,r)$. If we consider the 
time derivatives as small quantities and neglect the terms of the 
order $V^{2}$, the field equations are given by~\cite{HK76} 
and~\cite{her97}
\begin{align}
      e^{-\lambda }\left( \frac{\lambda ^{\prime }}{r}-\frac{1}{r^{2}}\right) +
      \frac{1}{r^{2}} &= 8\pi G \rho ^{({\rm eff})},  
      \label{stab1} \\
      e^{-\lambda }\left( \frac{\nu ^{\prime }}{r}+\frac{1}{r^{2}}\right) -
      \frac{1}{r^{2}} &= 8\pi G p_{r}^{({\rm eff})},
\end{align}
\begin{multline}
      \frac{1}{2}e^{-\lambda }\left(\nu''+\frac{\nu'^{2}}{2}+
      \frac{\nu'-\lambda'}{r}-\frac{\nu'}\lambda'{2}\right)\\ -
      \frac{1}{2}e^{-\lambda }\ddot{\lambda} = 8\pi G p_{\perp }^{({\rm eff})},
\end{multline}
\begin{align}
      e^{-\lambda /2}\frac{\dot{\lambda}}{r} =
      8\pi G\left[ \rho ^{({\rm eff})}+p_{r}^{({\rm eff})}\right] \dot{r},  
      \label{stab4}
\end{align}
where a dot means the derivative with respect to $t$.

We perturb the equilibrium solution
\begin{align}
      \rho^{({\rm eff})} &= \rho _{0}^{({\rm eff})} + 
      \delta \rho ^{({\rm eff})},\\ 
      p_{r}^{({\rm eff})} = p_{0r}^{({\rm eff})} +  
      \delta p_{r}^{({\rm eff})},\\
      p_{\perp }^{({\rm eff})} &= p_{0\perp }^{({\rm eff})} + 
      \delta p_{\perp }^{({\rm eff})},
\end{align}
\begin{align}
      \nu = \nu_{0} + \delta \nu ,\quad
      \lambda =\lambda_{0} + \delta \lambda ,
\end{align}
where the index $0$ refers to the equilibrium quantities. By
expanding Eqs.~(\ref{stab1})--(\ref{stab4}) to first order in the
perturbations we obtain~\citep{HK76}
\begin{align}
      8\pi G\delta p_{r}^{({\rm eff})}=\frac{e^{-\lambda_{0}}}{r^{2}}
      \left[r( \delta \nu' - \nu_{0}\delta\lambda') -\delta \lambda \right],
\end{align}
\begin{multline}
      8\pi G\delta p_{\perp }^{({\rm eff})} = \frac{1}{2}r
      \frac{\partial \delta p_{r}^{({\rm eff})}}{\partial r}+\delta p_{r}^{({\rm eff})}\\ +
      \frac{1}{4}\nu _{0}^{\prime }r\left[ \delta \rho^{({\rm eff})}+
      \delta p_{r}^{({\rm eff})}\right]\\ +
      \left[ \rho _{0}^{({\rm eff})}+p_{0r}^{({\rm eff})}\right] 
      \left[ \frac{1}{2} e^{\lambda _{0}}r\ddot{r}+\frac{1}{4}r\delta \nu'\right],
\end{multline}
\begin{align}
      8\pi G \delta \rho^{({\rm eff})}=\frac{1}{r^{2}}\frac{\partial}{\partial r}
      (re^{-\lambda _{0}}\delta \lambda),\\
      8\pi G \left[ \rho _{0}^{({\rm eff})} + p_{0r}^{({\rm eff})}\right]
      r\dot{r}=-e^{-\lambda_{0}} \delta \dot{\lambda}.
\end{align}

In the case of the Einstein cluster we have $p_{0r}^{({\rm eff})}=0$, and since 
the spherical symmetry of the perturbed system is preserved, we may also take 
$\delta p_{r}^{({\rm eff})}=0$. By assuming that all the perturbations are due 
to a small perturbation in $r=r(r,t) =r_{0}+\delta r(r,t)$, and after introducing 
the ansatz $\delta r(r,t) =e^{i\omega t}\xi(r)$ and denoting
$v(r) =e^{-i\omega t}r^{2}e^{-\nu /2}\delta r(r,t)$, the equation 
governing the radial pulsations of the Einstein cluster reduces to 
the simple algebraic condition
\begin{align}
      8\pi G \rho_{0}^{({\rm eff})} e^{\lambda _{0}-\nu _{0}}\left[
      \frac{e^{5\nu _{0}/2+\lambda _{0}/2}}{r^{2}}-\omega ^{2}\right] 
      v(r) = 0.
\end{align}
Since $v(r) \neq 0$, we obtain the condition 
\begin{align}
      \omega^{2}=e^{5\nu _{0}/2+\lambda _{0}/2}/r^{2}>0, \quad 
      \forall r\in \left[0,R\right].
\end{align} 
Stability means that the eigenfrequency $\omega$ of
the lowest mode is real and positive, $\omega^{2}>0$~\citep{HK76},
which taking into account the previous condition, is obviously
satisfied for the Einstein cluster. We therefore can conclude that
the Einstein cluster is stable against small radial perturbations
(a similar result was obtained by~\cite{Gi54} by using a different
method). The frequency of the radial oscillations of the cluster
is related to its radius $R$ and mass $M$ by the relation $\omega
^{2}=\left( 1-2GM/R\right) ^{2}/R^{2}$.

\subsection{Stability against non-radial perturbations}

For isotropic spherically symmetric systems the stability against
radial pulsations implies the stability against all small
adiabatic oscillations~\citep{Bi87,Pe}. For anisotropic models this need
not to be true since the anisotropy may serve as a source of instability
once the exact spherical symmetry is broken~\citep{HK76}. In the case of 
Einstein clusters the condition of the exact circular orbits 
implies the vanishing of the radial pressure. However, as far as this
condition is broken, and the trajectories of the particles are not
perfectly circular, a radial pressure may be generated, and this
pressure, together with the tangential pressure, may destabilise
the system. Therefore it is important to also analyse the stability of
the Einstein cluster under non-radial perturbations. Since the
general relativistic analysis of this problem is extremely
complicated and analytic solutions are difficult to be obtained,
we restrict ourselves to the Newtonian case and follow the
analysis given by~\cite{HK76}.

We assume that the perturbed Einstein cluster with non-spherical symmetry
has a density $\rho $, a radial pressure $p_{r}$ and a tangential pressure 
$p_{\perp }$, and that  generally $p_{r}\neq p_{\perp }$. As a
result of the perturbation, the Newtonian equations of motion of
the perturbed system, moving with velocity $\vec{v}$ under the action of 
the hydrodynamic force $\vec{F}$ in a gravitational field with potential 
$\Phi $ will be the anisotropic Euler equation, the continuity equation 
and the Poisson equation, given by
\begin{align}
      \rho \frac{d\vec{v}}{dt}=\vec{F}-\rho \nabla \Phi,
      \frac{\partial \rho }{\partial t}+\nabla \cdot(\rho\vec{v})=0,
\end{align}
and
\begin{align}
      \nabla^{2}\Phi = 4\pi G\rho,
\end{align}
respectively. In spherical coordinates $(r,\theta,\varphi)$ the components of 
the hydrodynamical force $F$ are
\begin{align}
      F_{r}=-\frac{\partial p_{r}}{\partial r}+\frac{2\Delta }{r},
      \nonumber \\
      F_{\theta }=-\frac{1}{r}\frac{\partial p_{\perp }}{\partial \theta },\quad
      F_{\varphi }=-\frac{1}{r\sin \theta }\frac{\partial p_{\perp }}{\partial \varphi }.
\end{align}

To close the system of equations we need an equation of state of
the radial pressure, which we assume to be of the barotropic form,
$p_{r}=p_{r}(\rho) $, and an equation of state for the anisotropy parameter $%
\Delta $, which we take as $\Delta =\beta \left( \rho \right) p_{r}$, with $%
\beta \left( \rho \right) $ an arbitrary function. We also denote
$\gamma =\left( \rho /p_{r}\right) \left( \partial p_{r}/\partial
\rho \right) $. In the following we will assume only axially symmetric 
perturbations, so that $\delta \varphi =0$.

By assuming that $\delta \vec{r}\left( \vec{r},t\right)
=e^{i\sigma t}\delta \vec{r}\left( \vec{r}\right) $, the
linearised equations of motion become
\begin{multline}
      \sigma ^{2}\delta r = \frac{\partial \delta \Phi }{\partial r}-
      \frac{\delta\rho }{\rho ^{2}}\frac{\partial p_{r}}{\partial r}+\frac{1}{\rho }
      \frac{\partial \delta p_{r}}{\partial r} \\ -
      \frac{2}{r\rho }\left( \delta \Delta -\frac{\delta r}{r}
      \Delta -\frac{\delta \rho }{\rho }\Delta \right),
      \label{nonr1}
\end{multline}
\begin{align}
      \sigma^{2}\delta \theta =\frac{1}{r}\frac{\partial }{\partial \theta }
      \left( \delta \Phi +\frac{\delta p_{\perp }}{\rho }\right),
\end{align}
\begin{align}
      \frac{\delta \rho }{\rho }+\frac{\delta r}{r}\frac{\partial \rho }{\partial r}+
      \frac{1}{r^{2}}\frac{\partial }{\partial r}\left( r^{2}\delta r\right) +
      \frac{1}{r\sin \theta }\frac{\partial }{\partial \theta }
      (r\sin \theta \delta \theta) = 0,
\end{align}
\begin{align}
      \nabla^{2}\delta \Phi = 4 \pi G\delta \rho,
\end{align}
respectively, where all the quantities without $\delta $ are
equilibrium quantities. We now express all the perturbations in 
terms of the Legendre polynomials $P_{l}(\cos\theta)$, so that
\begin{align}
      \delta \Phi = \delta \Phi ^{\ast }(r)P_{l}(\cos\theta),\quad
      \delta \rho = \delta \rho ^{\ast }P_{l}(\cos\theta),\quad 
      {\rm etc.}
\end{align}
Eq.~(\ref{nonr1}) thus becomes
\begin{multline}
      \label{eqmot}
      \sigma^{2}\delta r^{\ast } = \frac{\partial \delta \Phi ^{\ast}}{\partial r}-
      \frac{\delta \rho ^{\ast }}{\rho ^{2}}\frac{\partial p_{r}}{\partial r}+
      \frac{1}{\rho }\frac{\partial \delta p_{r}^{\ast }}{\partial r} \\ -
      \frac{2}{r\rho }\left( \delta \Delta ^{\ast }-\frac{\delta r^{\ast }}{r}
      \Delta -\frac{\delta \rho ^{\ast }}{\rho }\Delta \right) ,
\end{multline}
By introducing the new variables $\xi =\gamma p_{r}\delta \rho
^{\ast }/\rho $, $\varepsilon =\delta \Phi ^{\ast }$ and $\eta
=r^{2}\delta r^{\ast }$ we obtain the non-radial perturbation
equations for an anisotropic system as~\citep{HK76}
\begin{multline}
      \eta' = -\frac{p_{r}'}{\gamma p_{r}}+\frac{l(l+1)}{\sigma ^{2}}\varepsilon \\ -
      \left\{\frac{r^{2}}{\gamma p_{r}}-\frac{l(l+1)}{\sigma ^{2}\rho }
      \left[ \left( \beta +1\right) +\frac{\rho }{\gamma }\left( 
      \frac{\partial \beta }{\partial \rho }\right) \right] \right\} \xi,
      \label{nradf1}
\end{multline}
\begin{align}
      \varepsilon''=4\pi G\frac{\rho }{\gamma p_{r}}\xi -
      \frac{2}{r}\varepsilon'+\frac{l(l+1)}{r^{2}}\varepsilon,
      \label{nradf2}
\end{align}
\begin{multline}
       \xi' = \left( \sigma^{2}\rho + \frac{2\beta}{r^{2}}p_{r}\right) \frac{\eta }{r^{2}}-
       \rho \varepsilon' \\ -
       \frac{2}{r}\rho \left[ \frac{\beta }{\rho }\left( 1-\frac{1}{\gamma }\right) +
       \frac{1}{\gamma }\left( \frac{\partial \beta }{\partial \rho }\right) +
       \frac{1}{\gamma \rho }\frac{p_{r}'}{p_{r}}\right] \xi.
       \label{nradf3}
\end{multline}
Eqs.~(\ref{nradf1})--(\ref{nradf3}) must be solved together with
the boundary conditions $\xi(0) =\eta(0)=\varepsilon(0) =0$, $\varepsilon $ 
continuous across the boundary $R$ and $\Delta p_{r}^{\ast }\left( R\right)
=\left.\delta p_{r}^{\ast }\left( r\right) +p_{r}^{\prime }\delta
r^{\ast }(r)\right| _{r=R}=0$. Any solution satisfying these
conditions determines one eigenvalue $\sigma ^{2}$~\citep{HK76}.

We consider now the Einstein cluster as our equilibrium model.
Moreover, we will consider only quadrupole oscillations, thus
taking $l=2$. Since in this case $p_{r}=0$, it immediately follows 
that $\xi =0$. Hence Eq.~(\ref{nradf2}) becomes
\begin{align}
      \varepsilon''+\frac{2}{r}\varepsilon'-\frac{6}{r^{2}}\varepsilon = 0,
\end{align}
with the solution $\varepsilon(r) =C_{1}r^{2}$, with $C_{1}$ an arbitrary 
constant of integration. Eq. (\ref{nradf3}) gives $\varepsilon'=\sigma^{2}\eta /r^{2}$, 
and after taking the derivative of Eq.~(\ref{nradf1}) it follows that 
$\eta $ satisfies the second order differential equation
\begin{align}
      \eta'' - \frac{6}{r^{2}}\eta = 0,
\end{align}
with the solution $\eta(r) =C_{2}r^{3}$. From the
definition of $\eta $ we thus obtain $\delta r^{\ast }=C_{2}r$.
The continuity of the gravitational potential at the boundary
$r=R$ gives the ratio of the integration constants as
$C_{1}/C_{2}=GM/R^{3}$. In the case of the Einstein cluster the
condition $\Delta p_{r}^{\ast }(R) =0$ gives $\delta
p_{r}^{\ast }(R) =0$.

For the Einstein cluster $\Delta \approx \rho V_0^2/2$ and
$V_0^2\approx Gm(r)/r$, where $V_0={\rm constant}$ is the speed in
the constant velocity region. By estimating Eq.~(\ref{eqmot}) near
$r=R$ gives 
\begin{align}
      \left.\sigma^2(r)\right|_{r=R} = \left.2GM/R^3+Gm(r)/r^3\right|_{r=R} > 0. 
\end{align}
Since $\sigma ^2$ is always positive, it follows that the Einstein clusters 
are stable with respect to the quadrupole oscillations.

\section{Physical bounds for the Einstein clusters}
\label{sec4}

In realistic physical models for general relativistic matter distributions,
the anisotropy $\Delta$ should be finite, positive and should satisfy the
dominant energy condition (DEC) $\Delta \leq \rho $ and the strong energy
condition (SEC) $2\Delta \leq \rho $, see e.g.~\cite{BoHa06}. 
These conditions may be written together as $\Delta \leq n\rho $, 
where $n=1$ for DEC and $n=1/2$ for SEC, and they are automatically 
satisfied by the components of the energy-momentum tensor of the
Einstein cluster.

If the function $\Delta e^{\nu/2}/r$ is monotonically decreasing
for all $r$, then, as was proved rigorously in~\cite{BoHa05},
inside the anisotropic matter distribution the following
inequality always holds
\begin{align}
      \sqrt{1-\frac{2 G m(r)}{r} -\frac{\Lambda}{3}r^2} \geq \frac{1}{3}
      \left(1-\frac{\Lambda}{\langle\rho\rangle}\right)\frac{1}{1+f},
      \label{buchdahl}
\end{align}
where $f=f(m,r,\Lambda,\Delta)$, and we have already inserted that
$p_r=0$. The function $f$ explicitly reads
\begin{align}
      f = 2\frac{\Delta(r)}{\langle\rho\rangle}\left\{
      \frac{\arcsin\sqrt{2G\alpha(r)m(r)/r}}
      {\sqrt{2G\alpha(r)m(r)/r}}-1\right\},
\end{align}
where we have denoted
\begin{align}
      \alpha (r)=1+\frac{\Lambda r^{3}}{6 G m(r)},
\end{align}
and where $\langle \rho \rangle $ is the mean density of the matter
$\langle \rho \rangle =3m(r)/4\pi r^{3}$ inside the radius $r$.

For small values of the argument, we can use the series expansion
$\arcsin x /x \approx 1 + x^{2} / 6 +\ldots$, and obtain
\begin{align}
      f \approx \frac{2}{3} \frac{\Delta(r)}{\langle \rho \rangle }
      \frac{G \alpha(r) m(r)}{r}.
\end{align}

Since for the Einstein cluster the function $\Delta e^{\nu/2}/r =
1/r^{3-V_0^2}$ is monotonically decreasing, the above bound can be
applied to the constant velocity region. Let us estimate
Eq.~(\ref{buchdahl}) at the boundary $r=R$ of the cluster and
denote by $M=m(R)$ its total mass. By neglecting small terms
$(\Lambda r^2\ll 1, \Lambda V_0^2 \ll 1)$ the above bound reduces
to
\begin{align}
      \sqrt{1-\frac{2 G M}{R}} \geq \frac{1}{3}\left( 1 -
      \frac{\rho(R)}{3\langle\rho\rangle} V_0^2 \frac{G M}{R} \right).
      \label{condi}
\end{align}
Both terms containing $M/R$ can be replaced by the rotational velocity
in view of Eq.~(\ref{vel2}), where the cosmological constant is neglected
again. Hence, in terms of $V_{0}$, the condition given by Eq.~(\ref{condi})
can be reformulated as
\begin{align}
      V_{0}^{2} \leq \frac{4}{9} \left( 1 +
      \frac{4}{9}\frac{\rho(R)}{27\langle\rho_R\rangle }\right),
\end{align}
where $\langle\rho_R\rangle$ is the average density of the whole
cluster. On the other hand, by using the definition of the mean
density and Eqs.~(\ref{vel2}) and~(\ref {densit}) we have
\begin{align}
      \frac{\rho(R)}{\langle\rho_{R}\rangle}=\frac{1}{3}.
\end{align}
Hence, one obtains the following absolute bound for the rotational
velocity of the massive particles in the Einstein cluster
\begin{align}
      V_{0}^{2} \leq \frac{2}{3}\sqrt{1+\frac{2^2}{3^6}}.
      \label{condvel}
\end{align}

Therefore, we find the condition $V_{0} \leq 2/ 3 \approx 2 \times
10^{8} {\rm m/s}$. This result is consistent with the bound
obtained in~\cite{Gi54} from stability considerations, requiring
that the velocity of particles in stable circular orbits in
Einstein clusters must be smaller than half of the speed of light.
On the other hand, the Buchdahl limit for the mass-radius ratio
for the Einstein cluster (insert $G M/R$ for $V_0^2)$) gives no
physically relevant restriction since for typical galaxies $M/R
\ll 1$. The same analysis can in principle be repeated without
neglecting the contributions due to the cosmological constant and
the anisotropy in Eq.~(\ref{buchdahl}). However, the resulting
bound on the tangential velocity is only mildly effected and the
main result is unchanged.

Following the analysis of~\cite{BoHa06} it is natural to consider next
the behaviour of the invariant curvature scalars $r_0 = R$, $r_1 = R_{ab} R^{ab}$
and $r_2 = R_{abcd} R^{abcd}$ which are, in general, decreasing functions
with respect to the radius for regular relativistic matter distributions.
For the Einstein cluster, these invariants only yield the speed of light
as an upper bound on the velocity and no further condition emerges.

In the presence of the cosmological constant, the Buchdahl
inequality~(\ref{buchdahl}) not only leads to upper bounds on the
mass, but also implies the existence of a minimal
mass~\citep{BoHa05}. For Einstein clusters with radii $R_g$ in the
range $R_g\approx 10 \;{\rm kpc}-100 \;{\rm kpc}$ this yields
\begin{align}
      M_{\rm min} = \frac{4\pi}{3}\frac{\Lambda }{16\pi G}R_g^3
      \approx 4.5 \times 10^{5}-4.5\times 10^8 M_{\odot},
\end{align}
perfectly consistent with present mass estimates of galaxies. In
fact, this minimal mass is roughly of the order of the total
(virial) mass of the galaxies, and further supports that the
minimal mass due to $\Lambda$ is an important physical quantity in
an astrophysical context, see e.g.~\cite{Ba04,Ba05} and~\cite{Ba05b}.

\section{Light deflection and lensing by Einstein clusters}
\label{sec5}

One of the ways we could in principle test the possible existence
of Einstein clusters as astrophysical systems would be by studying
the light deflection in the cluster, and in particular by studying
the deflection of photons passing through the region where the
rotation curves are flat, with $V = V_0 = {\rm constant}$.

The metric coefficient $\exp{(\nu )}$ can be found from
Eq.~(\ref{nu}), and is given by
\begin{align}
      e^{\nu} = \left(\frac{r}{R_c}\right)^{2V_0^2},
\end{align}
where $R_c$ is a constant of integration. Its value can be found
by matching the metric tensor at the boundary $r=R$ with the standard
Schwarzschild-de Sitter metric. Thus we obtain
\begin{align}
      e^{\nu} = \left(1-2V_0^2-\frac{\Lambda}{3}R^2\right)
      \left(\frac{r}{R}\right)^{2V_0^2},
\end{align}

Let us consider a photon approaching a galaxy from far distance.
The bending of light by the gravitational field  results in a
deflection angle $\Delta \phi $ given by
\begin{align}
      \label{defl1}
      \Delta \phi =2\left| \phi \left( r_0\right) -\phi _{\infty}\right| -\pi ,
\end{align}
where $\phi_{\infty}$ is the incident direction and $r_0$ is the
coordinate radius of the closest approach to the centre of the
galaxy. Generally, one finds~\citep{BhKa03}
\begin{align}
      \phi \left( r_{0}\right) -\phi _{\infty } = \int_{r_{0}}^{\infty}
      \frac{e^{\lambda(r)/2}}{\sqrt{e^{\nu(r_{0}) -\nu(r)}\left(\frac{r}{r_{0}}\right)^{2}-1}}
      \frac{dr}{r}.
\end{align}
For the Einstein cluster in the constant velocity region we obtain
\begin{align}
      \phi(r_{0}) -\phi_{\infty}=\frac{
      \int_{r_{0}}^{\infty }\left[ \left( \frac{r}{r_{0}}\right)^{2(1-V_{0}^{2})}
      -1\right]^{-1/2}dr/r}{\sqrt{1-2V_{0}^{2}}}.
\end{align}

By introducing the new variable $\eta =r/r_{0}$ this leads to
\begin{align}
      \phi(r_{0}) -\phi_{\infty}=\frac{
      \int_{1}^{\infty }\left[\eta^{2(1-V_{0}^{2})}-1\right]^{-1/2}
      d\eta /\eta }{\sqrt{1-2V_{0}^{2}}},
\end{align}
and this integral can be evaluated exactly. Thus, we find the deflection angle
for the Einstein cluster as
\begin{align}
      \Delta \phi_{EC} =\frac{2\pi V_{0}^{2}}{\sqrt{1-2V_{0}^{2}}
      \left(1-V_{0}^{2}\right) }\approx 2\pi V_{0}^{2}.
\end{align}

In the standard approach to dark matter lensing it is assumed that the deflection
angle $\Delta\phi_{DM}$ is given by $\Delta\phi_{DM} = 4GM(r)/r$,
where $M(r)$ is the mass inside a radius $r$~\cite{Wa98}. In the constant velocity
region, the usual deflection angle therefore becomes
\begin{align}
      \Delta\phi_{DM} = 4 V_{0}^{2},
\end{align}
which is differs from the Einstein cluster deflection angle by a factor
of $\pi/2$.

In the generic case of a three-dimensional mass distribution, the density
$\rho(\vec{r})$ can be projected along the line of sight into the lens plane
to obtain the two-dimensional surface mass density distribution
$\Sigma(\vec{\xi})=\int_{0}^{D_{S}}\rho(\vec{r}) dz$, where $D_{S}$ is the
distance from the source to the observer and $\vec{\xi}$ is a two-dimensional
vector in the lens plane~\cite{Wa98}.

If one assumes for the galaxy lenses the singular isothermal
density sphere model, with density varying as $\rho(r)
=V_{0}^{2}/4\pi Gr^{2}$, the circularly symmetric mass
distribution is given by $\Sigma(\xi)=V_{0}^{2}/4G\xi $. Since
$M(\xi)=\int_{0}^{\xi}\Sigma(\xi') 2\pi \xi'd\xi'=\pi
V_{0}^{2}/2G\xi $, we find $\alpha_{DM}(\xi)=2\pi V_{0}^{2}$,
which is constant, that is, independent of the impact parameter
$\xi$~\citep{Wa98}. In the case of the Einstein cluster we obtain
\begin{align}
      \alpha_{EC}(\xi) = \frac{\pi \left( \pi /2+1\right) V_{0}^{2}}
      {\sqrt{1-\pi V_{0}^{2}}(1-V_{0}^{2})}
      \approx \pi \left( \frac{\pi }{2}+1\right) V_{0}^{2}.
\end{align}

Therefore, lensing effects can in principle discriminate between two different
dark models. For the Einstein cluster model in comparison with the isothermal
dark matter halo, we find
\begin{align}
      \frac{\alpha_{DM}}{\alpha_{EC}} =
      \frac{4}{\pi + 2} \approx 0.77.
\end{align}
This suggest that a galactic dark matter halo consisting of weakly
interacting massive particles in the form of an Einstein cluster
predicts slightly smaller gravitational lensing effects. Thus,
lensing seems to be a prime experimental tool to test different
dark matter models since they in general predict different
deflection angles.

\section{Conclusions}
\label{sec6}

The flattened galactic rotation curves and the absence of sufficient
luminous matter to explain them continue to pose a challenge to present 
day astrophysics. One would like to have a better understanding of some 
of the phenomena associated with them, like their universality and the 
very good correlation between the amount of dark matter and the luminous 
matter in galaxies.

In the present paper we have further developed an alternative view to the 
dark matter problem proposed by~\cite{La06}, namely,
the possibility that the dark matter is in the form of an Einstein
cluster of WIMPs. This assumption can explain the observed linearly
increasing mass profile outside galaxies, without introducing any
supplementary conditions. In fact, this model can reproduce any 
velocity profile~\cite{La06}. We analysed the stability of the 
Einstein clusters against both radial and non-radial perturbations, 
and we showed that the system is dynamically stable.

The Buchdahl bounds for the clusters have also been obtained, and
thus we showed that there is a maximal bound for the velocity
of the particles, and a minimum mass for the Einstein cluster.
This minimal mass is of the same order of as the total (virial) masses 
of the galaxies. The existence of the minimum mass is a direct 
consequence of the presence of a non-zero cosmological constant~\cite{BoHa05}.

All the relevant physical quantities have been expressed in terms of 
observable parameters (mass, radius and velocity dispersion). This allows
an in depth comparison of the predictions of the Einstein cluster model 
with observational results.

A possibility of observationally testing the viability of the
Einstein cluster as a dark matter model is via gravitational
lensing. We studied the lensing effect for Einstein clusters,
and compared it with the standard dark matter model, the singular
isothermal sphere. Even though the differences in the
deflection angles are small, a significant improvement in the
observational techniques may allow to discriminate between the
Einstein cluster and other dark matter models.

\section*{Acknowledgements}

We would like to thank Roy Maartens for his valuable comments. 
The work of CGB was supported by research grant
BO 2530/1-1 of the German Research Foundation (DFG). The work of
TH is supported by the RGC grant No. 7027/06P of the government of
the Hong Kong SAR.

\label{lastpage}


\begin{thebibliography}{99}

\bibitem[\protect\citeauthoryear{Balaguera-Antolinez et al.}{2005}]{Ba04} 
Balaguera-Antolinez A., B\"ohmer C.G., Nowakowski M., 2005, IJMP, D14, 1507

\bibitem[\protect\citeauthoryear{Balaguera-Antolinez and Nowakowski}{2005}]{Ba05b} 
Balaguera-Antolinez A., Nowakowski M., 2005, A\&AS, 441, 23

\bibitem[\protect\citeauthoryear{Balaguera-Antolinez et al.}{2006}]{Ba05} 
Balaguera-Antolinez A., B\"ohmer C.G., Nowakowski M., 2006, CQG, 23, 485

\bibitem[\protect\citeauthoryear{Banerjee and Som}{1981}]{BaSo81} 
Banerjee A., Som, M.M., 1981, Prog. Theor. Phys., 65, 1281

\bibitem[\protect\citeauthoryear{Binney and Tremaine}{1987}]{Bi87} 
Binney, J., Tremaine, S., 1987, {\it Galactic dynamics}, Princeton University Press, 
Princeton

\bibitem[\protect\citeauthoryear{B\"ohmer and Harko}{2005a}]{BoHa05a} 
B\"ohmer C.G., Harko T., 2005a, PRD, 71, 084026

\bibitem[\protect\citeauthoryear{B\"ohmer and Harko}{2005b}]{BoHa05} 
B\"ohmer C.G., Harko T., 2005b, PLB, 630, 73

\bibitem[\protect\citeauthoryear{B\"ohmer and Harko}{2006}]{BoHa06} 
B\"ohmer C.G., Harko T., 2006, CQG, 23, 6479

\bibitem[\protect\citeauthoryear{Comer and Katz}{1993}]{Co1} 
Comer G.L., Katz J., 1993, CQG, 10, 1751

\bibitem[\protect\citeauthoryear{Comer et al.}{1993}]{Co2} 
Comer G.L., Langlois D., Peter P., 1993, CQG, 10, L127

\bibitem[\protect\citeauthoryear{Einstein}{1939}]{ein} 
Einstein A., 1939, Annals Math., 40, 922

\bibitem[\protect\citeauthoryear{Gilbert}{1954}]{Gi54} 
Gilbert C., 1954, MNRAS, 114, 628

\bibitem[\protect\citeauthoryear{Herrera and Santos}{1997}]{her97} 
Herrera, L., Santos, N.O., 1997, Phys. Repts., 286, 53

\bibitem[\protect\citeauthoryear{Hillebrandt and Steinmetz}{1976}]{HK76} 
Hillebrandt, W., Steinmetz, K.O., 1976, A\&A, 53, 283.

\bibitem[\protect\citeauthoryear{Hledik et al.}{2007}]{Hl07}
Hledik, S., Stuchlik, Z., Mrazova,K., gr-qc/0701051

\bibitem[\protect\citeauthoryear{Hogan}{1973}]{Ho73} 
Hogan P., 1973, Proc. Royal Irish Acad., Sect.~A, 73, 91

\bibitem[\protect\citeauthoryear{Lake}{2006}]{La06} 
Lake K., 2006, gr-qc/0607057

\bibitem[\protect\citeauthoryear{Landau and Lifshitz}{1998}]{LaLi} 
Landau L.D., Lifshitz E.M., 1998, {\it The classical theory of fields}, 
Butterworth-Heinemann

\bibitem[\protect\citeauthoryear{Persic et al.}{1996}]{Pe}
Persic, M.,Salucci, P., Stel, F., 1996, MNRAS, 281, 27

\bibitem[\protect\citeauthoryear{Wambsganss}{1998}]{Wa98} 
Wambsganss J., 1998, Living Rev. Rel., 1, 12

\bibitem[\protect\citeauthoryear{Weinberg}{1972}]{BhKa03} 
Weinberg S., 1972, {\it Gravitation and Cosmology : Principles and Applications of the General Theory of Relativity}, Wiley \& Sons, New York

\end{thebibliography}
\end{document}